\documentclass[conference, letterpaper]{IEEEtran}
\IEEEoverridecommandlockouts

\usepackage{cite}
\usepackage{amsmath,amssymb,amsfonts}
\usepackage{url} 
\usepackage{algorithmic}
\usepackage{graphicx}
\usepackage{textcomp}
\usepackage{xcolor}
\usepackage{float}
\usepackage{url}
\usepackage{makecell}
\usepackage{booktabs}
\usepackage{multirow}
\def\BibTeX{{\rm B\kern-.05em{\sc i\kern-.025em b}\kern-.08em
    T\kern-.1667em\lower.7ex\hbox{E}\kern-.125emX}}
\begin{document}

\title{IceHorizon: A Dataset for Horizon Detection in Ice-Covered Maritime Environments and Comparative Evaluation of Detection Methods\\
\thanks{This work was supported by Mid Sweden University internal funding.}
}

\author{\IEEEauthorblockN{Alisa Pesotskaia}
\IEEEauthorblockA{\textit{Department of Computer and Electrical Engineering} \\
\textit{Mid Sweden University}, Sundsvall, Sweden \\
0009-0000-6247-279X}
\and
\IEEEauthorblockN{Emin Zerman}
\IEEEauthorblockA{\textit{Department of Computer and Electrical Engineering} \\
\textit{Mid Sweden University}, Sundsvall, Sweden \\
0000-0002-3210-8978}
}

\maketitle

\begin{abstract}
Horizon detection in images of ice-covered waters is a challenging problem for maritime navigation due to low contrast between water and sky, cluttered ice structures, and varying illumination conditions. This paper presents a comparative evaluation of six horizon detection algorithms, including four classical computer vision methods and two hybrid approaches combining deep learning with classical line detection. A new bespoke IceHorizon dataset consisting of 30 ship-based and 8 drone-based videos is used to evaluate detection accuracy, horizon coverage, and computational performance. The results show that hybrid methods achieve the highest accuracy and most reliable horizon estimates. In contrast, purely classical methods exhibit reduced robustness, particularly in visually ambiguous scenes. Performance on ship-based imagery was consistently higher than on drone-based imagery, indicating a strong dependency on acquisition characteristics. The created dataset and codes used in this study are made publicly available to support further research on this topic. The code is available at \url{https://github.com/allythe/HorizonDetection}. The dataset is available at \url{https://doi.org/10.5281/zenodo.20411867}.
\end{abstract}


\section{Introduction}
\label{sec:intro}
Horizon detection plays a central role in ensuring safe navigation or aircrafts and ships, especially in diverse and challenging environments~\cite{grelsson2016highly, cornall2006aircraft}. Increasing global maritime activity and the opening of new Arctic shipping routes due to climate change are driving the demand for advanced surveillance systems capable of operating in the challenging environments of ice-covered waters~\cite{chen2021perspectives}. 
The horizon, which in practice often manifests as a piecewise-linear contour rather than a perfect straight line, provides essential information for image stabilization on dynamic platforms \cite{schwendeman2015horizon, morris2007image}, camera auto-calibration \cite{zhang2012camera}, and platform attitude estimation for roll and pitch \cite{cornall2006aircraft, grelsson2016highly}. Accurate horizon estimation can also support camera–IMU extrinsic calibration and the detection of extreme vessel motions \cite{bovcon2018stereo}. 

Classical methods for horizon detection typically rely on traditional computer vision approaches, to estimate the horizon \cite{libe2012comparison}, such as edge detection, gradient analysis, and line fitting. While these methods can perform well under clear and simple conditions, they often struggle in the presence of noise, occlusions, reflections from ice or water surfaces, and complex lighting conditions.

\begin{figure}[t]
\centering
\includegraphics[width=0.43\textwidth]{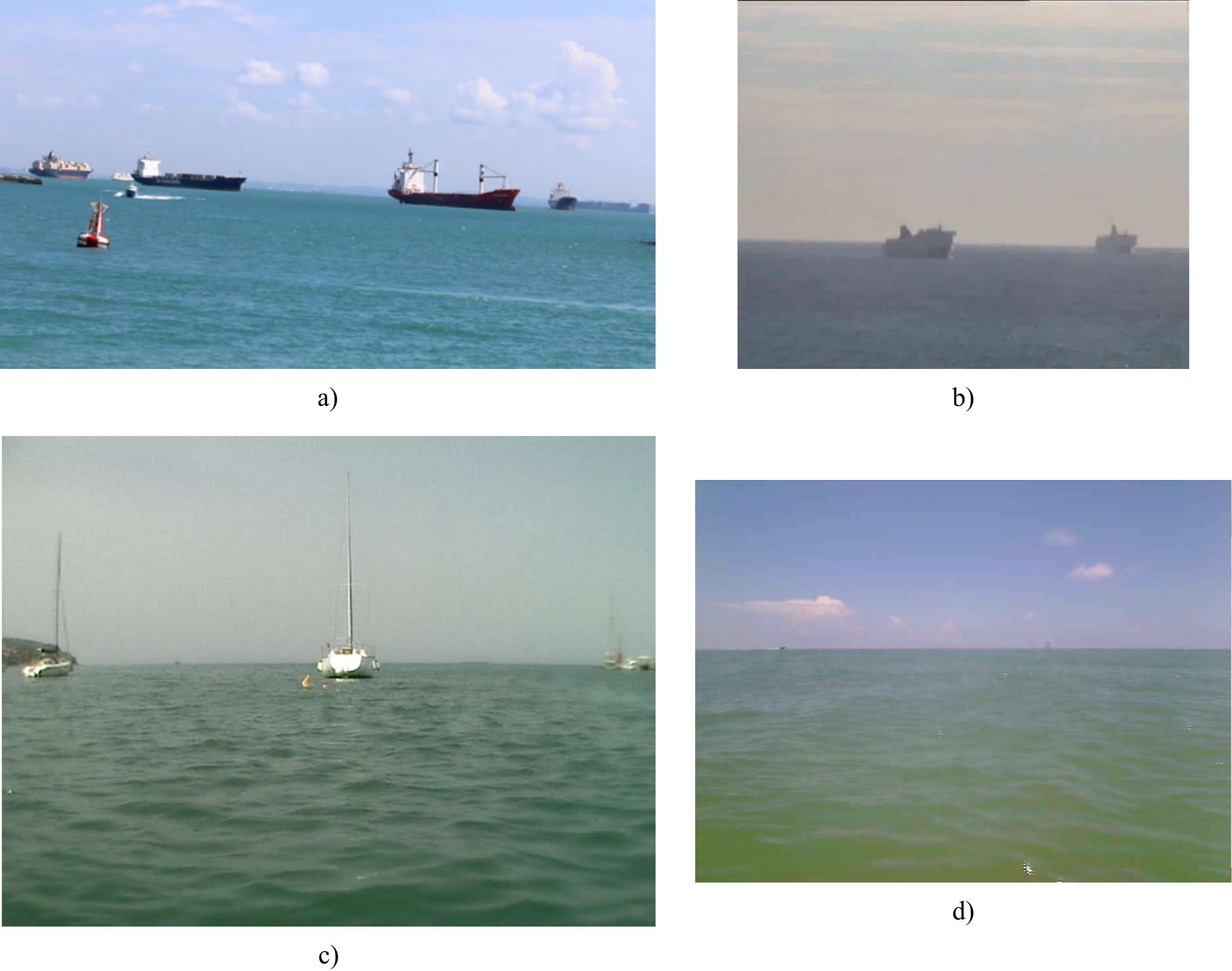}
\caption{Sample frames from existing open-water maritime datasets: (a) Singapore Maritime Dataset~\cite{prasad2017video}, (b) MarDCT~\cite{bloisi2015argos}, (c) Marine Obstacle Detection dataset~\cite{kristan2015fast}, (d) Buoy dataset~\cite{fefilatyev2006horizon}. These datasets primarily represent open-water conditions and lack examples of ice-covered marine environments.}
\label{fig:datasets}
\vspace{-3mm}
\end{figure}

\begin{figure*}[!t]
    \centering
    \includegraphics[width=0.95\textwidth]{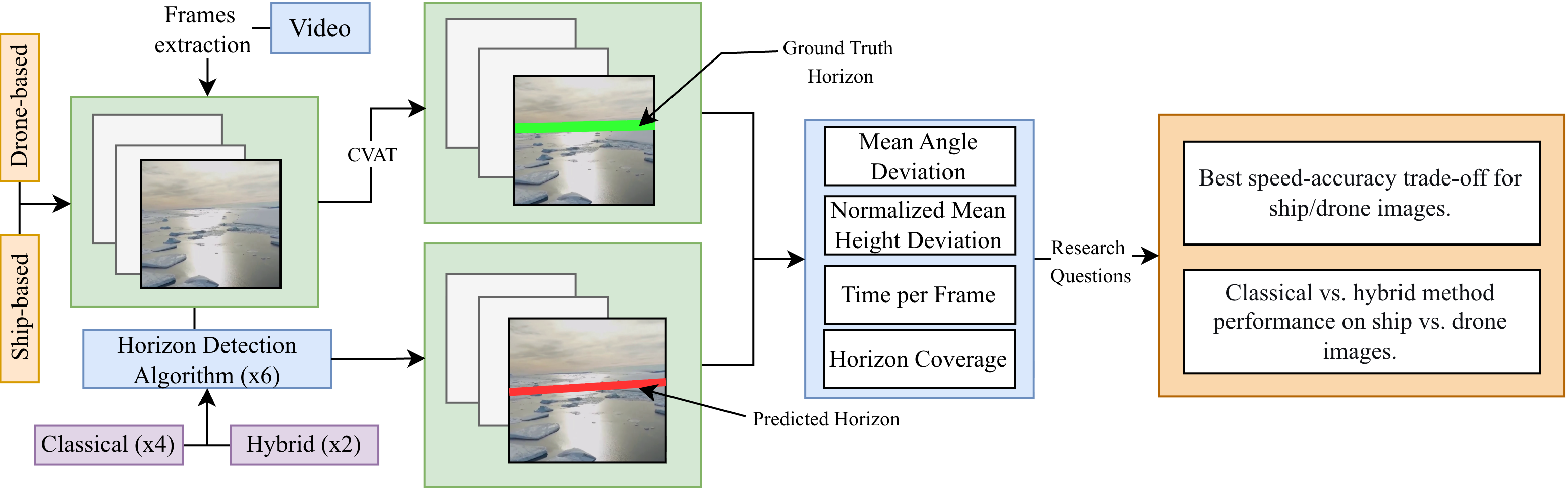}
    \caption{Overview of the horizon detection evaluation pipeline for ice-covered waters. Input videos are split into frames, with ground truth horizons manually annotated using CVAT. Six algorithms (H-HC, H-MED, H-LSC, H-COV-LUM, DexiNed, Segment) predict horizon lines for each frame. Performance is quantified using mean angle deviation (MAD) and normalized mean height deviation (NMHD), predicted horizon coverage, and time spent per frame, enabling comparative analysis to address the research questions.}
    \label{fig:overview}
    \vspace{-3mm}
\end{figure*}

Deep learning methods~\cite{sandru2023horizon}, particularly convolutional neural networks (CNNs), have recently emerged as promising alternatives for maritime perception tasks, offering improved robustness by learning semantic representations of the sky and sea. However, their performance strongly depends on the diversity of the training data and architectural optimizations~\cite{zardoua2023survey}. 
Existing maritime datasets (see Fig.~\ref{fig:datasets}), including the Singapore Maritime Dataset~\cite{prasad2017video}, MarDCT~\cite{bloisi2015argos}, Marine Obstacle wDetection~\cite{kristan2015fast}, and the Buoy dataset~\cite{fefilatyev2006horizon}, primarily focus on open-water environments and therefore do not capture the visual challenges of ice-covered waters.  More recently, datasets such as the Tangier Maritime Dataset~\cite{zardoua2021fast} and several deep-learning-based horizon detection approaches have been proposed, however, these methods are still largely designed for ice-free maritime environments. For example, Chen et al.~\cite{chen2024innovation} introduced a hybrid pipeline that combines HED and Canny edge detection with a ConvNeXt-based edge classification network and RANSAC line fitting for sea-sky line detection in maritime imagery. 


The limited data diversity and the need for further algorithmic refinement to handle ice-induced visual artifacts were also highlighted in a recent CNN-based horizon detection study~\cite{sandru2023horizon}. Existing methods and datasets do not adequately address horizon detection in ice-covered maritime environments, where visual ambiguity, reflections, and rapidly changing weather conditions introduce new challenges.

This paper aims to provide new insights into horizon detection in ice-covered waters  by creating a new annotated dataset of ship and drone-based videos recorded in ice-covered waters and performing a comparative evaluation of classical and hybrid methods that combine deep learning with classical approaches. 
The overview of the horizon detection and evaluation pipeline is shown in Fig.~\ref{fig:overview}. 
In particular, this paper has two specific contributions:
\begin{itemize}
    \item A new IceHorizon dataset on icy-water horizon detection including 38 videos and ground truth annotations, which is made publicly available to serve as a reference for future research; and
    \item A benchmark of traditional and hybrid horizon detection methods in the newly created icy-water horizon detection IceHorizon dataset, complete with statistical analyses.
\end{itemize}




\section{Selected Horizon Detection Methods}
\label{sec:horizonDetectionMethods}

The existing methods on horizon detection use either traditional computer vision approaches~\cite{libe2012comparison} or deep-learning-based approaches~\cite{sandru2023horizon}. While classical approaches use low-level features, deep-learning-based approaches utilize a data-oriented approach through feature extraction specific for the horizon detection task. Hybrid methods, on the other hand, can leverage both the robustness of the traditional approaches and a semantic understanding of the scene.


In a recent survey on maritime horizon detection, Zardoua et al. \cite{zardoua2023survey} concluded that classical methods suffer from inherent trade-offs between robustness and real-time performance, with edge-based approaches being vulnerable to competitive lines and regional-based methods failing under class variation and non-homogeneous regions. They emphasized that learning-based approaches, particularly neural networks, show superior discrimination capability but require significant architectural optimization for real-time maritime applications. Similarly, Petković et al. \cite{petkovic2020overview} concluded that while classical methods remain more practical for real-time maritime applications due to their computational efficiency, methods based on neural networks demonstrate superior robustness and accuracy.

In this section, we describe how the selected horizon detection methods work.

\subsection{Classical methods}
Classical horizon detection methods rely on low-level image features such as gradients, edges, and regional statistics. These approaches typically assume that the horizon corresponds to a strong linear discontinuity between the sky and the sea. While efficient and widely used, their performance can degrade under challenging conditions such as reflections, occlusions, and low-contrast ice-covered waters. The classical methods evaluated in this work are summarized below.

\textbf{H-HC}: This edge‐detection and Hough transform–based algorithm first pre-processes the image using erosion or a low-pass filter to suppress noise. It then applies a Canny edge detector \cite{canny1986edge} and uses the Hough transform \cite{duda1972hough} to identify line candidates. The longest detected line is selected as the horizon. To improve robustness, a dilation step is introduced prior to the Hough transform to strengthen and connect edge segments.

\textbf{H-LSC}: The edge-detection and least-squares calibration–based algorithm pre-processes the image using erosion or a low-pass filter, identifies the strongest edge in each image column, and then fits a line through these edge points using a least-squares estimator.

\textbf{H-MED}: Median filtering and linear regression based algorithm algorithm follows the same overall procedure as H-LSC but differs in how edges are defined. Instead of using standard edge detection based on the absolute intensity differences between adjacent pixels, H-MED computes edge strength as the difference of local medians. 

\textbf{H-COV-LUM}: The regional covariance–based algorithm searches for an optimal partition of the image into two regions, sky and sea, using a candidate line that represents the horizon. The optimal horizon is selected by evaluating an objective function based on the determinants and traces of the covariance matrices computed for the two regions.

\subsection{Hybrid Methods}

Two hybrid methods are implemented for horizon detection to investigate the utility of merging the learning-based models with traditional approaches.

\textbf{Segment}: The first hybrid method uses a pre-trained UPERNet-ConvNeXt-Small~\cite{xiao2018unified, liu2022convnet} model for semantic segmentation. 
The CNN-based ConvNeXt-Small backbone extracts features to generate a segmentation mask of a set of types. After that, all segments which are quantified as water are put in the segmentation mask.  The H-HC algorithm is then applied to the segmentation mask to detect the longest line, which is considered the horizon. Processing the segmentation mask instead of the original image helps to reduce spurious edges caused by ice, waves, or clouds.

\textbf{DexiNed}: The second hybrid method employs the pre-trained edge detection CNN DexiNed \cite{soria2023dexined}. 
It produces an edge map of the image, and the Hough transform is applied to detect the longest line as the horizon. By using DexiNed’s learned edge features, this approach similarly reduces noise from irrelevant edges while directly leveraging a CNN trained for precise edge detection.

\section{Methodology}
\label{sec:experimentalSetup}


The selected horizon detection methods were compared in an experiment.
A new IceHorizon dataset with ice-covered water videos was created to make this comparison possible\footnote{The IceHorizon dataset is accessible on Zenodo: \url{https://doi.org/10.5281/zenodo.20411867}}.

\begin{figure}[!t]
\centerline{\includegraphics[width=0.46\textwidth]{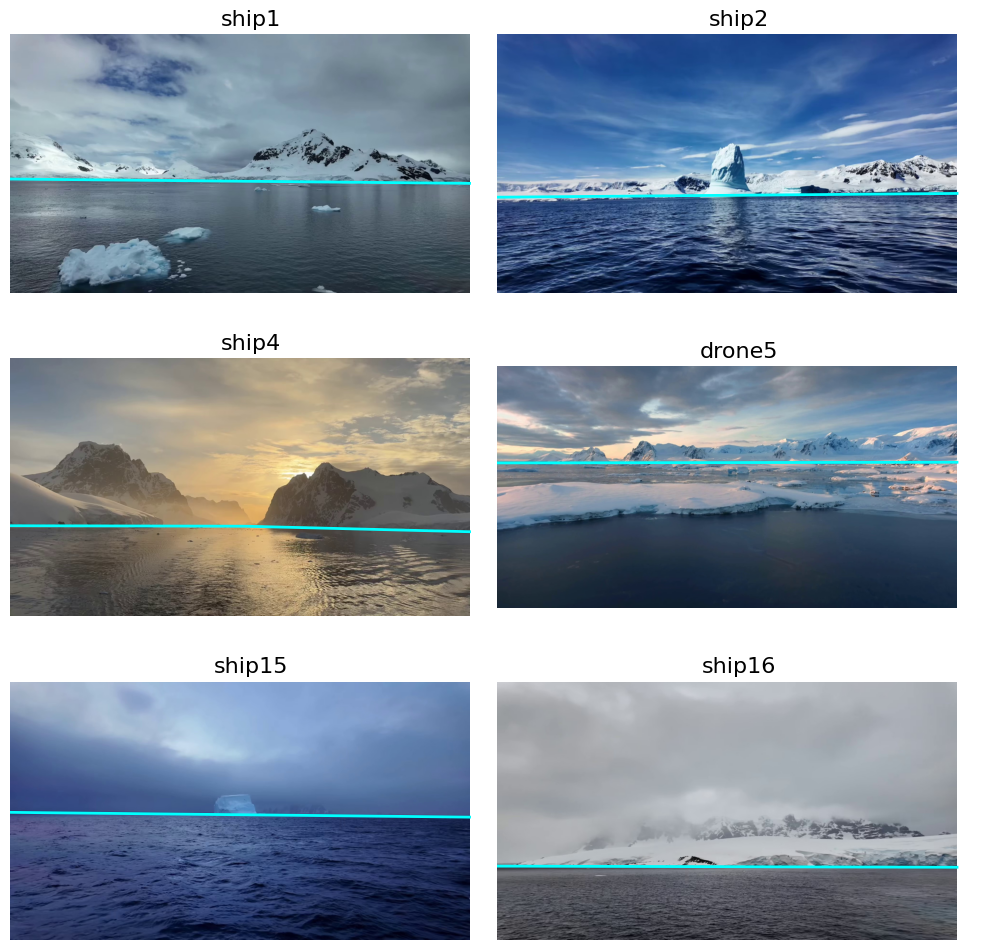}}
\caption{Representative frames from the IceHorizon dataset with ground-truth horizon annotations. }
\label{fig:sample_images}
\end{figure}

\begin{table}[bp]
\centering
\caption{Image resolutions of used videos.}\label{tab:videos} 
\begin{tabular}{lcccccc}
\toprule

\textbf{Video} & \cite{bo2024antarctica} & \cite{auteri2020arctic}  & \cite{wandering2023antarctica} & \cite{fluent2018island} & \cite{free2024antarctica} & \cite{civil2018weddell}\\
\midrule
\textbf{Ship}  & 1-22 & 23 & 24 & -   & - & 25-30\\
\textbf{Drone} & -    & -  & -  & 1-4 & 5 & 6-8 \\
\midrule
\textbf{Height}& 1080 & 720  & 1080 & 720  & 676  & 1080\\
\textbf{Width} & 1920 & 1280 & 1920 & 1280 & 1280 & 1920
\\
\bottomrule
\end{tabular}
\end{table}

\subsection{Dataset}

We created a new IceHorizon dataset to evaluate the horizon detection methods, which consists of frames extracted from six YouTube videos released under a Creative Commons Attribution license \cite{bo2024antarctica, auteri2020arctic, wandering2023antarctica, fluent2018island, free2024antarctica, civil2018weddell}. The first three videos primarily depict ice-covered waters from a ship-based perspective, while the remaining three are captured predominantly from a drone viewpoint. These videos were further segmented into 38 shorter clips, each maintaining a relatively consistent viewpoint. The horizon line in each frame was manually annotated using the Computer Vision Annotation Tool (CVAT)~\cite{sekachev2020cvat}. 
A summary of the video resolutions, the extracted frames, and representative examples with annotated horizons are provided in Table~\ref{tab:videos}, Table~\ref{tab:video_frames}, and  Figure~\ref{fig:sample_images}, respectively.

\begin{table}[t]
\centering
\caption{Number of frames extracted from each video sequence.}
\label{tab:video_frames}
\begin{tabular}{lrlrlr}
\toprule
\textbf{Video} & \multicolumn{1}{c|}{\textbf{Frames}} &
\textbf{Video} & \multicolumn{1}{c|}{\textbf{Frames}} &
\textbf{Video} & \textbf{Frames} \\
\midrule
Ship1  & 105  & Ship11 & 1031 & Ship21 & 159  \\
Ship2  & 83   & Ship12 & 207  & Ship22 & 125 \\
Ship3  & 86   & Ship13 & 342  & Ship23 & 724 \\
Ship4  & 48   & Ship14 & 452  & Ship24 & 2046 \\
Ship5  & 98   & Ship15 & 199  & Ship25 & 213 \\
Ship6  & 152  & Ship16 & 338  & Ship26 & 178 \\
Ship7  & 631  & Ship17 & 152  & Ship27 & 154 \\
Ship8  & 412  & Ship18 & 563  & Ship28 & 134 \\
Ship9  & 561  & Ship19 & 144  & Ship29 & 192 \\
Ship10 & 201  & Ship20 & 64  & Ship30 & 211 \\
\midrule
Drone1 & 57  & Drone4 & 461  & Drone7 & 182 \\
Drone2 & 203   & Drone5 & 927  & Drone8 & 143 \\
Drone3 & 251  & Drone6 & 131  &        &     \\
\midrule
\textbf{\makecell{Ship \\total}}  & \multicolumn{1}{c|}{\textbf{10005}} & 
\textbf{\makecell{Drone \\total}} & \multicolumn{1}{c|}{\textbf{2355}} &
\textbf{\makecell{Grand \\total}} & \textbf{12360} \\
\bottomrule
\end{tabular}
\end{table}

\subsection{Selected metrics for evaluation}

Horizon detection accuracy and performance were quantified using Normalized Mean Height Deviation (NMHD), Mean Angle Deviation (MAD)~\cite{gershikov2013horizon}, computational performance, and predicted horizon coverage. 

NMHD measures the average vertical difference between the detected and ground-truth horizons, normalized by the image height, and is defined as:
\begin{equation}
    \mathrm{NMHD} = \frac{1}{W H} \sum_{i=1}^{W} \left| h_i - \hat{h}_i \right|,
\end{equation}
where $h_i$ and $\hat{h}_i$ denote the ground-truth and detected horizon heights at column $i$, $W$ is the image width in pixels, and $H$ is the image height.

MAD quantifies the average angular difference between the detected and true horizon orientations and can be expressed in terms of the horizon angles \(\theta\) as:
\begin{equation}
    \mathrm{MAD} = \frac{1}{W} \sum_{i=1}^{W} \left| \hat{\theta}_i - \theta_i \right|,
\end{equation}
where \(\theta_i\) and \(\hat{\theta}_i\) denote the ground-truth and detected horizon angles at column \(i\).

It should be noted that neither NMHD nor MAD can be reliably computed from only two horizon points, since the ground-truth horizon may be represented as a piecewise-linear curve rather than a single straight line.

Computational performance is measured by the processing time (in seconds) per frame. 

PHC calculates the completeness of the detected horizon line. It is calculated as the ratio of the number of columns where the horizon is detected to the total number of columns where the horizon should appear:
\begin{equation}
    \mathrm{PHC} = \frac{N_{\mathrm{detected}}}{N_{\mathrm{truth}}},
\end{equation}

where $N_{\mathrm{detected}}$ is the count of columns with a valid detected 
horizon and $N_{\mathrm{truth}}$ is the count of columns where the 
ground-truth horizon is defined.

\begin{figure}[t]
\centerline{\includegraphics[width=0.5\textwidth]{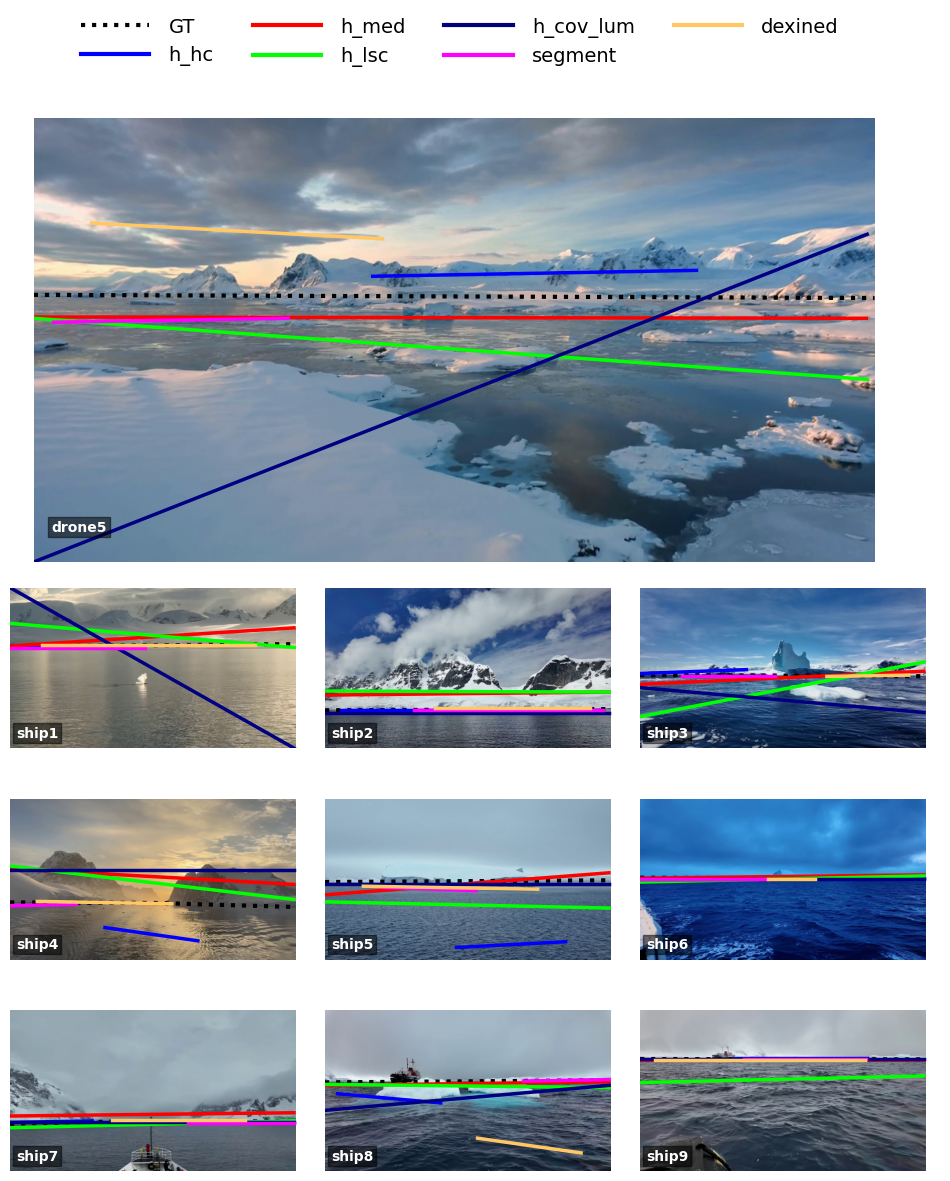}}
\caption{Representative frames from several video category showing horizon detection results. Ground truth annotations (black dashed lines) are compared with predictions from six methods: four classical (H-HC, H-MED, H-LSC, H-COV-LUM) and two hybrid (DexiNed, Segment).}
\label{fig:results_img}
\end{figure}

\subsection{Implementation Details}
All experiments were conducted on Google Colab using a Python 3 runtime. Classical computer vision algorithms were executed on the CPU with 12.7GB of system RAM, while hybrid methods were run on a T4 GPU with 15GB of GPU memory. The available disk space was approximately 112GB. This setup allowed efficient processing of the dataset and reproducible evaluation of algorithm performance.

\begin{figure}[htp]
\centerline{\includegraphics[width=0.95\columnwidth]{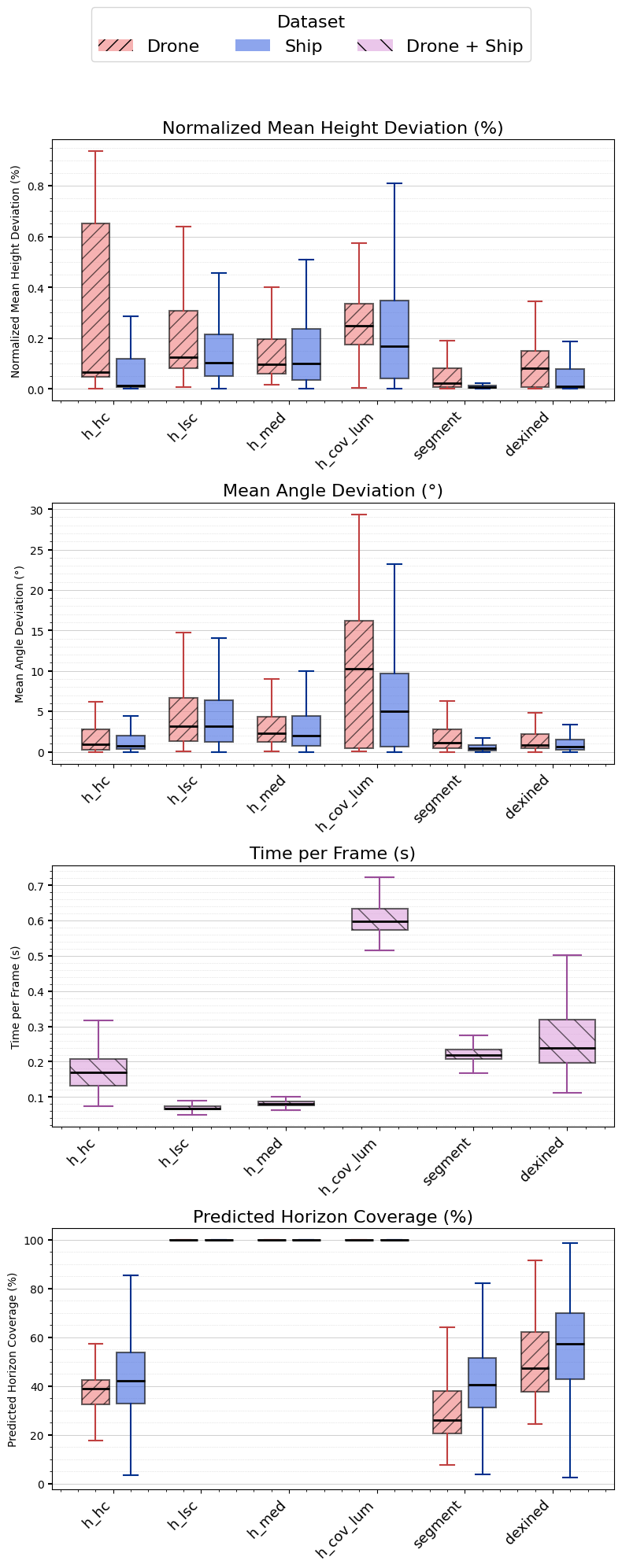}}
\caption{Boxplots comparing performance metrics for classical and hybrid horizon detection methods: Normalized Mean Height Deviation (NMHD), Mean Angle Deviation (MAD), Processing Time per Frame, and Predicted horizon Coverage. Results are shown for drone-based and ship-based images across all six methods. }
\label{fig:results}
\end{figure}

\section{Results}
\label{sec:results}

\begin{table}[]
    \centering
    \caption{Performance metrics, reported as "Mean (Std)" for each algorithm on ship and drone-based datasets. \textbf{Bold} indicates the best score and \textit{italic} indicates the second best score. \vspace{2mm}}
    \label{tab:metrics}
    \begin{tabular}{llccc}
        \toprule
          & \textbf{Method} & \textbf{NMHD} & \textbf{MAD} & \textbf{PHC} \\ \midrule
         \parbox[t]{2mm}{\multirow{6}{*}{\rotatebox[origin=c]{90}{\textbf{Ship}}}}
           & H-HC      & \textit{0.07 (0.11)} & 3.13 (8.03) & 44.7 (16.3) \\ 
           & H-LSC     & 0.14 (0.11) & 4.14 (3.50) & \textbf{100 (0)} \\
           & H-Med     & 0.14 (0.13) & 3.00 (2.99) & \textbf{100 (0)} \\
           & H-Cov-Lum & 0.22 (0.20) & 6.26 (6.67) & \textbf{100 (0)} \\
           & Segment   & \textbf{0.03 (0.09)} & \textbf{0.87 (2.40)} & 41.9 (16.3) \\
           & DexiNed   & 0.10 (0.17) & \textit{2.84 (8.79)} & \textit{56.0 (17.8)} \\ \midrule
         \parbox[t]{2mm}{\multirow{6}{*}{\rotatebox[origin=c]{90}{\textbf{Drone}}}}
           & H-HC      & 0.29 (0.31) & 3.77 (8.31) & 38.4 (11.7) \\ 
           & H-LSC     & 0.19 (0.15) & 5.29 (6.78) & \textbf{100 (0)} \\
           & H-Med     & 0.13 (0.10) & 3.16 (2.65) & \textbf{100 (0)} \\
           & H-Cov-Lum & 0.26 (0.13) & 10.4 (9.54) & \textbf{100 (0)} \\
           & Segment   & \textbf{0.08 (0.13)} & \textit{2.51 (4.44)} & 30.5 (14.0) \\
           & DexiNed   & \textit{0.10 (0.12)} & \textbf{1.95 (3.05)} & \textit{50.2 (14.3)} \\ 
         \bottomrule
    \end{tabular}
\end{table}

After running horizon detection methods on the proposed IceHorizon dataset, the evaluation metrics were computed.  
Sample results showing predicted and ground truth horizons are presented in Figure~\ref{fig:results_img}. Boxplots of the quality metrics are shown in Figure~\ref{fig:results}, with separate representations for drone and ship-based datasets for NMHD, MAD, predicted horizon coverage, and time spent per frame. 

Numerical results are reported in Table~\ref{tab:metrics}: hybrid methods provide up to x2.3-3.5 accuracy increase for ship-based videos and x1.6 accuracy increase  for drone-based videos against the best traditional horizon detection methods (H-HC and H-Med, respectively). Results also show that horizon coverage significantly reduces for the hybrid methods, which yields a trade-off with accuracy results.

As the results are not normally distributed, pairwise Mann–Whitney U tests with Holm-Bonferroni correction were conducted to identify the best-performing algorithms.

\textbf{NMHD Analysis}:
For the drone and ship categories, the NMHD of \textit{Segment} was significantly smaller than that of all other algorithms, with all pairwise Mann--Whitney U tests (Holm-corrected) yielding $p < 0.001$. \textit{Dexined} ranked second overall and was not significantly smaller only when compared with \textit{Segment}.

\textbf{MAD Analysis}:
For the drone category, the MAD of \textit{DexiNed} was significantly smaller than that of all other algorithms except \textit{H-HC}, while \textit{Segment} and \textit{H-HC} were not significantly smaller than \textit{DexiNed} or each other.
For the ship category, \textit{Segment} was significantly smaller than all other methods, while \textit{DexiNed} ranked second, with its values being significantly smaller than those of all methods except \textit{Segment}.

\textbf{Time per Frame Analysis}: 
For the whole IceHorizon dataset, H-LSC was the best-performing method, with Time per Frame being significantly smaller than that of all other methods. H-MED ranked second and was not significantly smaller only than H-LSC. 

\textbf{Predicted Horizon Coverage}: 
For the drone category, the PHC of H-LSC and H-COV-LUM was significantly greater than that of all other algorithms except for each other, with all pairwise Mann-Whitney U tests (Holm-corrected) yielding $p < 0.001$. H-MED was the runner-up, being not significantly greater than only H-LSC and H-COV-LUM.  
For the ship category, the situation was the same.

Overall, as in the NMHD and MAD analyses for drone and ship categories, hybrid methods Segment and Dexined are the optimal choices, yielding the smallest errors across both metrics.

Regarding time per frame and predicted horizon coverage, the classical methods H-LSC and H-MED demonstrate the strongest performance; however, their NMHD and MAD scores are not competitive with those of H-HC, Segment, or DexiNed. Among the top two algorithms, DexiNed attains the highest predicted coverage. Since the horizon line can be interpolated across the entire image, and given that it is nearly straight for most samples, predicted coverage should be treated as a supplementary metric rather than a primary one. 

Differences between the datasets follow a consistent pattern: both NMHD and MAD are lower for the ship category, whereas the drone category exhibits higher median error values. This performance gap can be attributed to greater scene complexity in the drone category, as discussed in Section~\ref{sec:discussion}. 


\section{Discussion}
\label{sec:discussion}

The 
Segment and DexiNed horizon detection methods achieved the best performance on ship-based and drone-based videos, respectively. 
All algorithms performed better on ship-based videos than on drone-based videos, despite the higher vantage point offered by the drone views, due to the increased scene complexity present in the latter. Both classical and hybrid horizon detection methods achieved substantially lower NMHD and MAD errors on ship-based videos compared to drone-based videos. This performance gap can be attributed to several characteristics of the drone-based subsets, including horizons that are partially obscured by fog, mountains, and large ice formations, which introduce strong distractor edges. In addition, some drone videos exhibit 
slightly curved horizon resulting from the high-altitude viewpoint and lens characteristics. These factors increase ambiguity and reduce the reliability of classical and hybrid methods when applied to drone-based imagery.

Overall, a more controlled comparison would benefit from capturing the same scenes from both a ship and a drone, ensuring that observed performance differences reflect viewpoint effects rather than scene variability. Additionally, expanding the dataset with a larger number of drone-based videos is necessary to improve statistical reliability and better assess algorithm performance under diverse high-altitude conditions. Finally, it should be noted that all evaluated algorithms are restricted to predicting a straight horizon line, which limits their ability to accurately model curved horizons commonly observed in high-altitude drone imagery.

\section{Conclusion}

In this study, we conducted a comprehensive evaluation of six horizon detection methods on a new IceHorizon dataset comprising 10 005 ship-based images and 2 355 drone-based images of ice-covered waters. 
The results show that
the hybrid methods achieved the best performance on both types of categories (i.e., ship and drone). This confirms that hybrid approaches offer a favorable trade-off between detection accuracy and computational efficiency. 
All selected horizon detection methods performed better on ship-based videos than on drone-based videos. This reflects the increased complexity and ambiguity inherent in the drone videos, which contained challenging conditions such as fog, ice formations, mountains, and curved horizons.

Future work should focus on expanding the drone category to improve statistical reliability and creating paired drone and ship-based videos of the same scenes, enabling controlled comparisons that isolate the effect of viewpoint on horizon detection performance.

\bibliographystyle{IEEEbib}
\bibliography{refs}

\end{document}